\documentclass[aps,prl,reprint,groupedaddress,notitlepage]{revtex4-1}
\usepackage{graphicx}
\graphicspath{{figures/}}
\usepackage{dcolumn}
\usepackage{bm}
\usepackage{cancel}
\usepackage{xcolor}
\usepackage{amsmath}
\usepackage{amsfonts}
\graphicspath{{figures/}}

\usepackage{hyperref}
\usepackage{amscd}
\usepackage{placeins}
\usepackage{chngcntr}
\counterwithin*{equation}{section}
\begin{document}
\title{Nitrogen-Vacancy Ensemble Magnetometry Based on Pump Absorption}
\author{Sepehr Ahmadi, Haitham A.R. El-Ella, Adam M. Wojciechowski,\\Tobias Gehring, J{\o}rn B. Hansen, Alexander Huck, and Ulrik L. Andersen}
\affiliation{Department of Physics, Technical University of Denmark, 2800 Kongens Lyngby, Denmark}

\begin{abstract}
We demonstrate magnetic field sensing using an ensemble of nitrogen-vacancy centers by recording the variation in the pump-light absorption due to the spin-polarization dependence of the total ground state population. Using a \mbox{532 nm} pump laser, we measure the absorption of native nitrogen-vacancy centers in a chemical vapor deposited diamond placed in a resonant optical cavity. For a laser pump power of \mbox{0.4 W} and a cavity finesse of 45, we obtain a noise floor of \mbox{$\sim$ 100 nT/$\sqrt{\textrm{Hz}}$} spanning a bandwidth up to \mbox{125 Hz}. We project a photon shot-noise-limited sensitivity of \mbox{$\sim$ 1 pT/$\sqrt{\textrm{Hz}}$} by optimizing the nitrogen-vacancy concentration and the detection method.   
\end{abstract}
\maketitle

\section{Introduction}
The nitrogen-vacancy (NV) center in diamond is currently one of the most studied and anticipated platforms for high spatial-resolution sensing of magnetic fields \cite{Balasubramanian2008,Rondin2014,Schirhagl2014}, electric fields \cite{Dolde2011} and temperature \cite{Kucsko2013,Neumann2013} at ambient conditions. Several novel applications using diamond sensors are currently being developed in the fields of neuroscience \cite{Hall2012,Barry2016}, cellular biology \cite{Steinert2013,Glenn2015}, nanoscale magnetic resonance microscopy \cite{Lovchinsky2016}, paleomagnetism \cite{Fu2014}, and microelectronics \cite{Kolkowitz2015,Jakobi2016}.   

Many magnetometer schemes using NV centers are based on recording the change in the detected fluorescence level upon a shift of the electron spin precession frequency due to a change of an external magnetic field \cite{Clevenson2015a,Glenn2015,Barry2016,Sepehr2017}. The fluorescence contribution of an ensemble of NV centers to the signal increases the optically detected magnetic resonance (ODMR) amplitude and hence boosts the sensitivity by $\sqrt{N}$, where $N$ is the number of NV centers \cite{Taylor2008}.
However, the high refractive index of diamond (\mbox{$\sim$ 2.4}) together with the near uniform emission of NV center ensembles trap most of the generated fluorescence due to total internal reflection. This limits the collection efficiency and, thus, the smallest detectable magnetic field change. To increase the fluorescence collection from a diamond, several techniques have been demonstrated such as fabricating a solid immersion lens \cite{Hadden2010}, side-collection detection \cite{LeSage2012}, employing a silver mirror \cite{Israelsen2014}, using a dielectric optical antenna \cite{Riedel2014}, emission into fabricated nanopillar waveguides \cite{Momenzadeh2015} and employing a parabolic lens \cite{Wolf2015}. Alternatively, magnetic fields can also be sensed by observing the change in the shelving-state infrared absorption \cite{Jensen2014}, or the change in fluorescence when transitioning through the ground state level anti-crossing of the NV center \cite{Wickenbrock2016}. 

In this article, we report on a new measurement technique for NV ensemble magnetometry, which is based on monitoring the spin-dependent absorption of the pump field. Using the absorption detected magnetic resonance (ADMR) measurement technique in conjunction with a cavity resonant with the pump field, we fully circumvent challenges associated with inefficient collection of fluorescence, by detecting the absorption through the transmitted cavity mode. We demonstrate a NV ensemble magnetometer for low-frequency magnetic field sensing with a measured noise floor of \mbox{$\sim$ 100 nT/$\sqrt{\textrm{Hz}}$} spanning a bandwidth up to \mbox{125 Hz}. Intriguingly, using the reflection of an impedance-matched cavity and a diamond crystal with an optimized NV concentration, we project an estimated sensitivity of \mbox{$\sim$ 1 pT/$\sqrt{\textrm{Hz}}$}.

\section{Absorption detected magnetic resonance}
\begin{figure}
	\includegraphics[scale = 0.31]{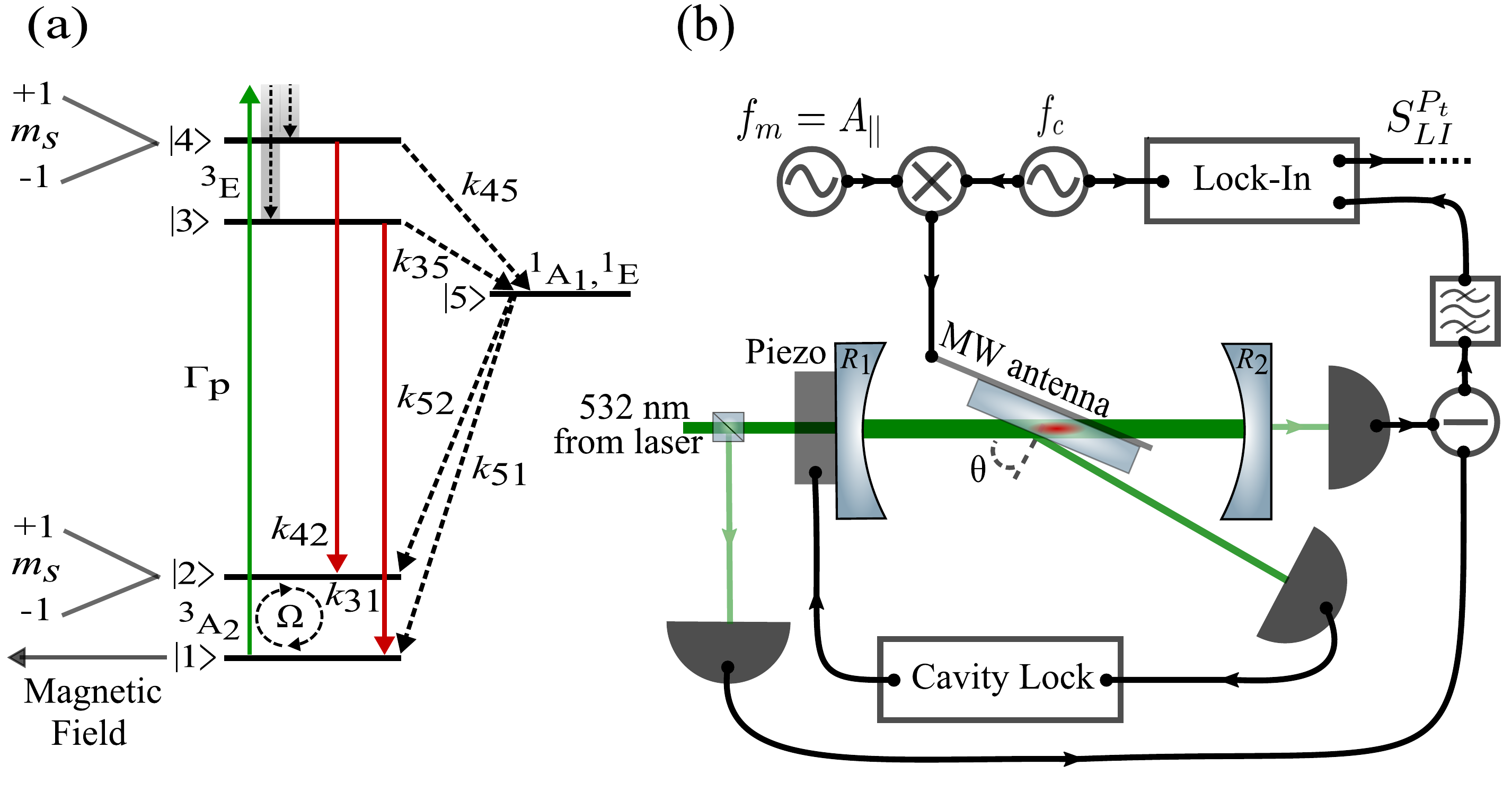}
	\caption{(a) Summary of the NV center energy levels and transitions between them.
Green laser light excites the NV center with a rate $\Gamma_p$ to a quasi-continuous vibronic state which decays quickly to the optical excited states.
The decay between two states is shown by $k_{ab}$, and $\Omega$ corresponds to the Rabi frequency of the MW drive.
Presence of a magnetic field lifts the degeneracy of \mbox{$m_s=\pm 1$} proportional to 2$\gamma_e B_{\text{NV}}$.
Non-radiative transitions are shown by dashed arrows. (b) Schematic of our experimental setup to perform ADMR measurements through the cavity transmission (see the main text and the supplemental material for further experimental details).}
	\label{Cavity}
\end{figure}

The electronic level structure of the NV defect is summarized in \mbox{Fig. \ref{Cavity}(a)}. It consists of a $^3$A$_2$ spin-triplet ground state, a $^3$E spin-triplet excited state, and a \mbox{$^1$A$_1$ $\leftrightarrow$ $^1$E} shelving state. Pumping with a \mbox{532 nm} laser results in an excitation above the zero phonon line, which decays on a picosecond timescale \cite{Huxter2013} to the $^3$E excited states by non-radiative transitions. Moreover, there exists a non-radiative decay path through the shelving state which is more probable for \mbox{$m_s=\pm 1$} of the excited state $|$4$\rangle$.
Continuous optical pumping depopulates the \mbox{$m_s=\pm 1$} spin sublevel and accumulates the population in \mbox{$m_s=0$}.
The zero-field splitting of the ground state levels $|$1$\rangle$ and $|$2$\rangle$ is \mbox{$\sim$ 2.87 GHz} at room temperature, making the transition between these levels accessible using microwave (MW) fields. The presence of a local magnetic field lifts the degeneracy of \mbox{$m_s=\pm$1} with a splitting proportional to 2$\gamma_e B_{\text{NV}}$, where \mbox{$\gamma_e$ = 2.8 GHz/T} is the gyromagnetic ratio of the electron spin and $B_{\text{NV}}$ corresponds to the magnetic field projection along the NV symmetry axis. A change in the external magnetic field hence results in a detectable shift in the electron spin resonance frequency of the ODMR or the ADMR spectrum, respectively.
The continuous-wave sensitivity of the spin resonances to small changes of an external magnetic field is proportional to max[$\frac{d}{d_w}S]^{-1}$, where $\frac{d}{d_w}$ is the derivative with respect to the MW frequency $\omega/2\pi$ of the ADMR signal $S$. Using a cavity around the diamond host crystal, a change in $S$ can be detected by a measurement of the remaining pump light either transmitted through or reflected off the cavity. Intriguingly, by appropriately tailoring the impedance of the cavity it is possible to obtain a unity contrast in the reflected light power, which in turn may lead to a sensitivity in the \mbox{pT/$\sqrt{\textrm{Hz}}$} range.

\section{Experiment}
We use the native $^{14}$NV$^-$ concentration of an off-the-shelf single-crystal diamond grown by chemical vapor deposition. A schematic of the experimental setup is shown in \mbox{Fig. \ref{Cavity}(b)}. The optical cavity consists of two concave mirrors with a \mbox{10 cm} radius of curvature set in a confocal configuration, resulting in a minimum beam waist of \mbox{92 $\mu$m} with a Rayleigh length of \mbox{$\sim$ 50 mm}. The mirrors have the measured reflectivities of \mbox{$R_1$ = 94.8$\%$ $\pm$ 0.1$\%$} and \mbox{$R_2$ = 99.8$\%$ $\pm$ 0.1$\%$} at the pump wavelength of \mbox{532 nm}.
With the diamond rotated at its Brewster angle \mbox{($\theta\simeq$ 67$^{\circ}$)}, the round-trip beam path in the diamond is \mbox{$l=$ 2 $\times$ 1.3 mm} and the estimated excitation volume is \mbox{$\sim$ 3.5 $\times$ 10$^{-2}$ mm$^3$}, accounting for the standing wave and the transverse beam profile.
The finesse of a cavity is defined by \mbox{$F=\pi\sqrt{\rho}/(1-\rho)$}, where \mbox{$\rho=\sqrt{R_1R_2e^{-\alpha}}$} corresponds to the cumulative round-trip loss product and $\alpha$ is the propagation loss coefficient. In the absence of the diamond, the finesse solely depends on the product of the mirror reflectivities $R_1R_2$ and is calculated as \mbox{$F = 113.4 \pm 4.4$}, which is confirmed by the measured finesse of \mbox{$F = 114 \pm 0.1$}. Incorporating the diamond into the cavity reduces the finesse to \mbox{$F = 45.1 \pm 0.1$}, which indicates that all the effective loss in the loaded cavity can be attributable solely to losses occurring through the diamond. The corresponding cumulative round-trip loss of the loaded cavity shows that the cavity is slightly under-coupled. The propagation loss can be decomposed to \mbox{$\alpha=\alpha_{abs}l+\alpha_r$}, in which $\alpha_{abs}$ is the absorption loss coefficient and $\alpha_{r}$ is attributed to all other loss channels such as surface-based absorption, scattering losses, and birefringence losses.
The total fraction of reflected light from the diamond to intra-cavity power was measured as \mbox{$\sim$ 0.006}, of which approximately 80$\%$ was $s$-polarized light.
This translates to an absorption loss coefficient of \mbox{$\alpha_{abs}\sim$ 0.0301 mm$^{-1}$,} taking \mbox{$\alpha_{r}\sim$ 0.006}.
With an independent measurement using a confocal microscope, we determined the NV$^-$ concentration, [NV$^-$], to be \mbox{$\sim$ 2.9 $\times$ 10$^{10}$ mm$^{-3}$} (\mbox{$\sim$ 0.16 ppb}) corresponding to \mbox{$\sim$ 10$^{9}$ NV$^-$} centers within the excitation volume. Considering the absorption cross section of a single $^{14}$NV$^-$ at \mbox{532 nm} (\mbox{$\sigma_{\text{NV}}=$ 3.1 $\times$ 10$^{-15}$ mm$^2$} \cite{Wee2007}), a NV related absorption loss coefficient of \mbox{$\alpha^{\text{NV}}_{abs}\sim$ 9 $\times$ 10$^{-5}$ mm$^{-1}$} is obtained. Hence, in our diamond sample most of the propagation loss is attributed to non-NV loss channels.
Using the NV absorption loss coefficient, we estimate the ratio between the excitation rate and the intra-cavity power \mbox{$\epsilon=\Gamma_p /P_{cav}\sim$ 75 kHz/W}, where the intra-cavity and incident powers are linked through \mbox{$ P_{cav} = P_{in}(1-R_1)/|1-\sqrt{R_1R_2e^{-\alpha}}|^2$}.

\begin{figure}
	\includegraphics[scale = 0.31]{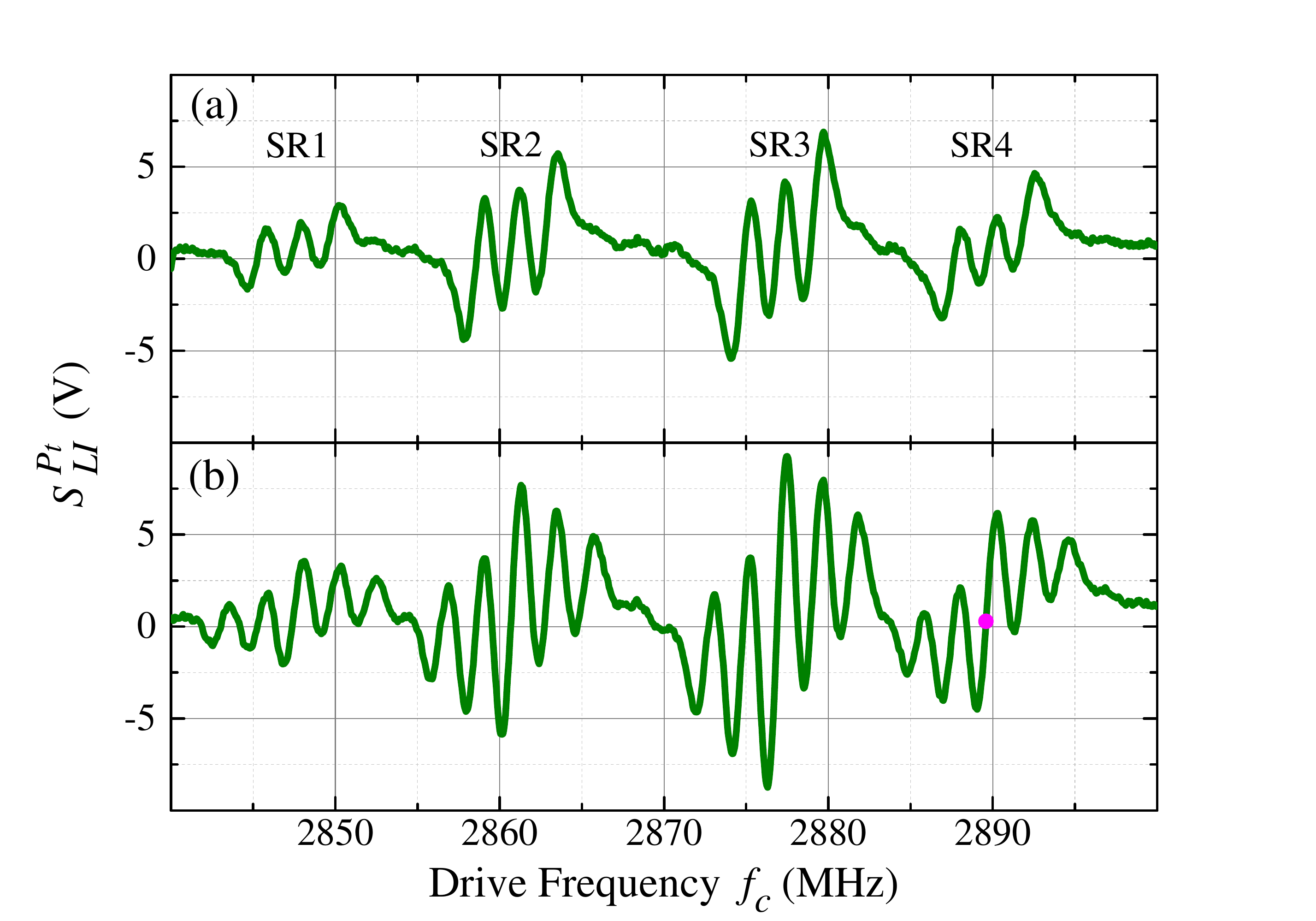}
	\caption{Measured frequency-modulated ADMR spectrum using (a) single-frequency excitation and (b) three-frequency excitation. SR1 and SR4 correspond to the electron spin resonances of single crystallographic orientation of NVs, while SR2 and SR3 correspond to the electron spin resonances of the other three crystallographic orientations. The purple dot in (b) indicates the point that is most sensitive to small changes in the magnetic field.}
	\label{ODMR}
\end{figure}
\subsection{Spectrum}
We performed ADMR measurements by recording the remaining pump-light transmitted through the diamond loaded cavity while sweeping the MW drive frequency across the spin resonance. To reduce the technical noise level in our measurement, we tapped off some laser light before the cavity, recorded it with a second photodetector and subtracted the two photocurrents, as indicated in \mbox{Fig. \ref{Cavity}(b)}. In order to remove low frequency technical noise, we applied lock-in detection with a frequency modulated MW drive, directly yielding $S^{P_t}_{LI}$ at the output, where $P_t$ indicates the transmitted power through the cavity and $LI$ refers to lock-in (further experimental details can be found in the supplemental material). A typical frequency modulated ADMR spectrum is presented in \mbox{Fig. \ref{ODMR}(a)}. In these measurements, a static magnetic field was aligned along the [111] axis, resulting in the outermost electron spin resonances (SR1,SR4), while the inner peaks (SR2,SR3) correspond to the electron spin resonances of the other three crystallographic orientations.
The three-peak feature of the ADMR spectrum in \mbox{Fig. \ref{ODMR}(a)} is a consequence of the hyperfine interaction between the NV electron spin and the intrinsic $^{14}$N nuclear spin with a coupling constant of \mbox{$A_{||}=2.16$ MHz} \cite{Smeltzer2009}.
To enhance max[$\frac{d}{d_w}(S^{P_t}_{LI}$)], we excited all three $^{14}$N hyperfine transitions simultaneously by mixing the modulation frequency $f_c$ with a \mbox{$f_m$ = $A_{||}$} signal. The three-frequency excitation results in five peaks for each electron spin resonance, as shown by the measured spectrum in \mbox{Fig. \ref{ODMR}(b)}.

\begin{figure}
	\includegraphics[scale = 0.3]{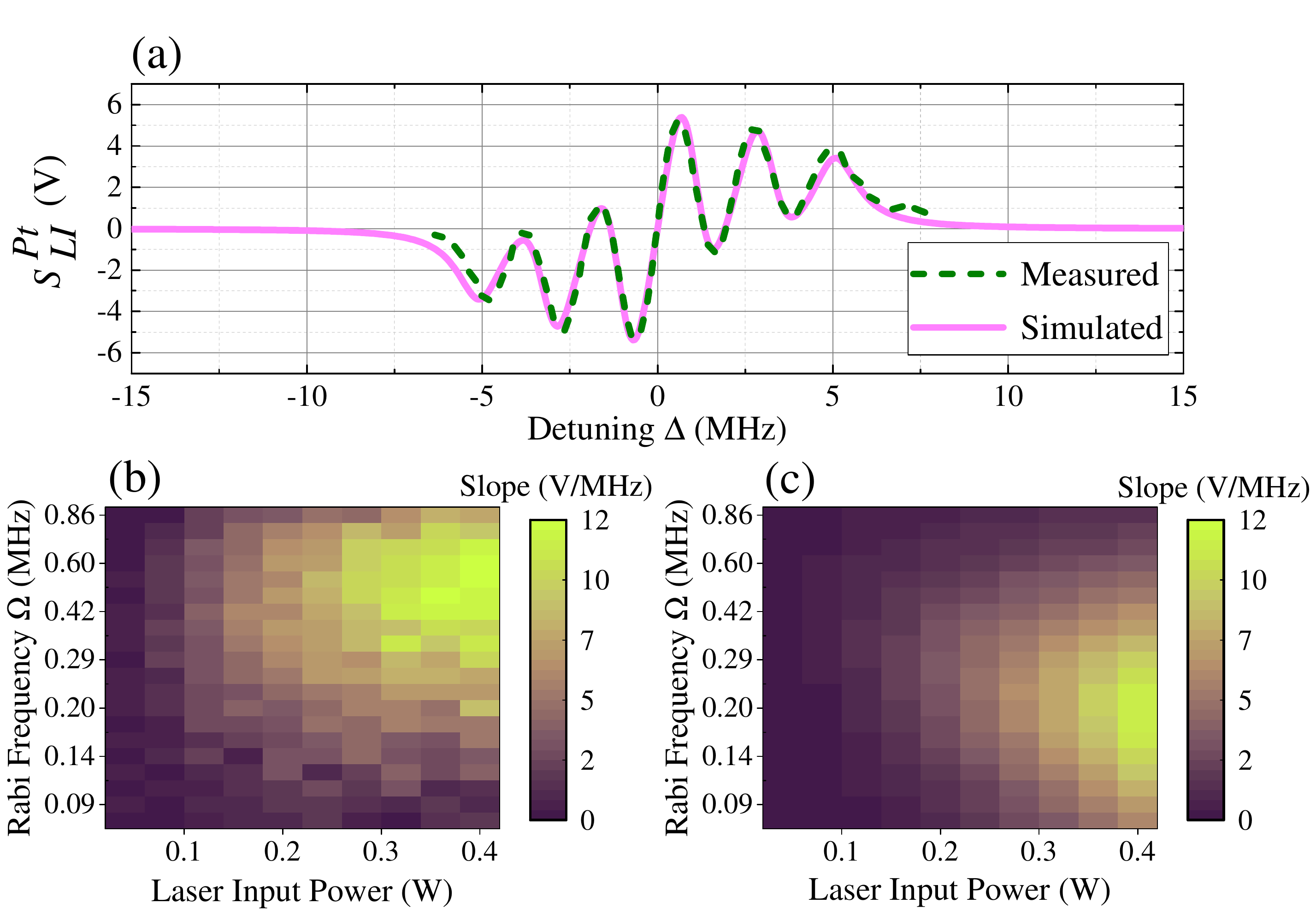}
	\caption{(a) Simulated and measured frequency-modulated ADMR spectra using three-frequency excitation. The used parameters are \mbox{$P_{in}$ = 0.4 W}, \mbox{$\Omega$ = 0.3 MHz}, \mbox{$R_1$ = 94.8 $\%$}, \mbox{$R_2$ = 99.8 $\%$}, \mbox{$\alpha^0_{abs}$ = 0.0781}, \mbox{$\alpha_r$ = 0.006}, \mbox{[NV$^-$] = 0.16 ppb}, \mbox{$l$ = 2$\times$1.3 mm}, \mbox{$\epsilon$ = 75 kHz/W}, \mbox{$\gamma^*_2$ = 1/3 MHz}, \mbox{$\gamma_1$ = 0.182 kHz}, and \mbox{$GV_0$ = 65 $\times$ 10$^6$ V}. (b) Measured and (c) simulated slopes of three-excitation, frequency-modulated ADMR spectra at \mbox{$\Delta=0$} as a function of $P_{in}$ and $\Omega$.
The maximum measured slope in (b) is obtained for \mbox{$P_{in}$ = 0.4 W} and \mbox{$\Omega\sim$ 0.3 MHz}.}
	\label{map}
\end{figure}

\begin{figure*}
	\includegraphics[scale = 0.3]{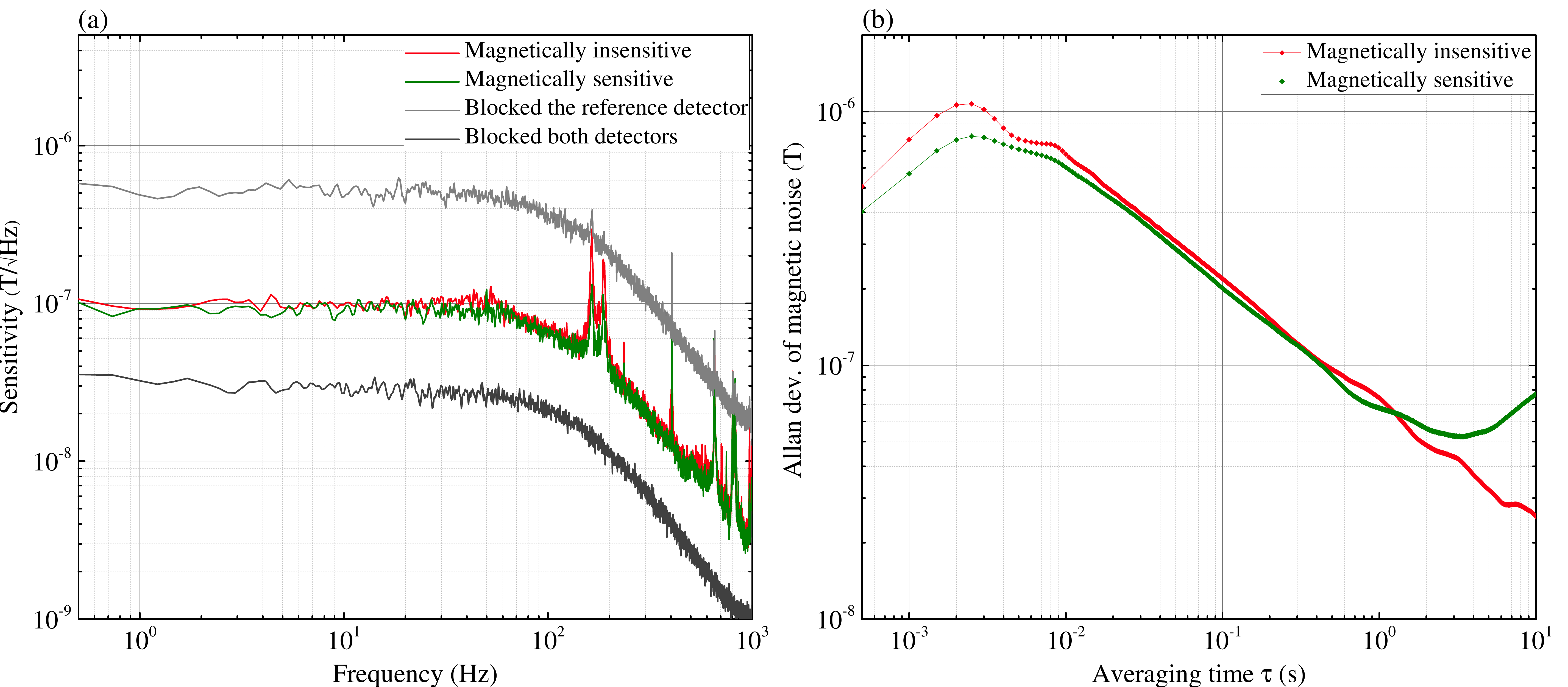}
	\caption{(a) Measurements of the magnetic noise spectral density: when the MW drive is set on the maximum slope of the frequency modulated ADMR (corresponding to the purple dot in \mbox{Fig. \ref{ODMR}(b)} - magnetically sensitive), when the MW drive is far from any spin resonance (magnetically insensitive), when we do not cancel out the correlated laser noise (blocked the reference detector), and the noise floor of the lock-in and blocked detectors for the same gain setting. (b) Measurements of the Allan deviation of magnetic noise of the traces in (a). The drop with the slope of -1/2 identifies the white noise in the system. For the magnetically sensitive trace, there is a minimum at \mbox{$\sim$ 3.3 s} which increases at higher averaging time due to thermal or mechanical drift in the system. The Allan deviation was calculated using the overlapping method.}
	\label{Pulsed}
\end{figure*}

\subsection{Model}
An ADMR spectrum $S_{LI}$ may be obtained either by recording the pump beam reflected from the cavity, $S^{P_r}_{LI}$, or transmitted through the cavity, $S^{P_t}_{LI}$, as a function of the applied MW frequency, and may be modeled using a set of optical Bloch equations considering the five electronic levels and the transitions summarized in \mbox{Fig. \ref{Cavity}(a)} \cite{Haitham2017}. The steady-state level populations $\rho^{ss}$ are then obtained as a function of Rabi frequency $\Omega$, optical excitation rate $\Gamma_p$, and MW detuning $\Delta$ from the spin \mbox{$m_s=0 \leftrightarrow m_s=\pm 1$} transition.
The cavity reflection or transmission itself is a function of loss inside the cavity which is dominated by the absorption in diamond, while the NV absorption in diamond depends on the NV ensemble ground state spin population. Applying a resonant MW field (\mbox{$\Delta=0$}) increases the population in the shelving state $|5\rangle$, which possesses a longer lifetime (\mbox{$ > 150$ ns} \cite{Robledo2011a, Acosta2010a}) than the $^3$E excited states, and hence, a lower average population remains in the ground states $|1\rangle$ and $|2\rangle$ to absorb the pump photons. Ultimately, the resonant MW field decreases the optical loss inside the cavity which can be monitored through the light transmitted or reflected from the cavity.
The steady-state population of the optical ground state can be written as:
\begin{equation}
\rho^{ss}_g(\Omega,\Gamma_p,\Delta) = \rho^{ss}_{11} + \rho^{ss}_{22},
\end{equation}
where $\rho^{ss}_{11}$ and $\rho^{ss}_{22}$ are the steady-state population of $|$1$\rangle$ and $|$2$\rangle$, respectively.
As the absorption of a NV ensemble directly depends on $\rho^{ss}_g$, a change in the propagation loss as a function of [NV$^-$] can be described as:
\begin{equation}
\alpha(\Omega,\Gamma_p,\Delta,[\text{NV}^-])=\alpha^0_{abs}+[\text{NV}^-]\sigma_{\text{NV}}l\rho^{ss}_g+\alpha_r,
\end{equation}
where $\alpha^0_{abs}$ is the loss coefficient attributed to non-NV absorption. As pump absorption in our sample is dominated by non-NV related processes, the absorption-based spin contrast $C_{\text{ADMR}}$ related to the fraction $\alpha^{\text{NV}}_{abs}/\alpha$ is on the order of $10^{-6}$ when monitoring the absorption through the cavity transmission. The steady-state cavity outputs as a function of MW detuning are then reformulated in terms of transmitted and reflected powers:
\begin{equation}
\frac{P_t}{P_{in}} = \frac{T_1T_2e^{-\alpha(\Omega,\Gamma_p,\Delta,[\text{NV}^-])}}{|1-\sqrt{R_1R_2e^{-\alpha(\Omega,\Gamma_p,\Delta,[\text{NV}^-])}}|^2},
\label{Ptra}
\end{equation}
\begin{equation}
\frac{P_r}{P_{in}} = \frac{(R_1-\sqrt{R_1R_2e^{-\alpha(\Omega,\Gamma_p,\Delta,[\text{NV}^-])}})^2}{R_1|1-\sqrt{R_1R_2e^{-\alpha(\Omega,\Gamma_p,\Delta,[\text{NV}^-])}}|^2},
\label{Pref}
\end{equation}
where $P_{in}$ is the laser input power to the cavity, $T_1$ and $T_2$ are the transmissions of the first and the second mirror, respectively, and we assume \mbox{$R_i+T_i=1$}. For the sake of simplicity, the intra-cavity excitation rate (\mbox{$\Gamma_p=\epsilon P_{cav}$}) is calculated in terms of the input power and the propagation loss when no MW field is applied (\mbox{$\Omega = 0$}, \mbox{$\Delta\rho^{ss}_g$ = 1}).
The lock-in signal $S^{P_i}_{LI}$ can be described as a function of detuning between the carrier frequency $f_c$ and the resonance frequency $f_0$ (\mbox{$\Delta=f_c - f_0$}), and the modulation depth $\delta$ through:
\begin{equation}
\begin{split}
S^{P_i}_{LI}(\Delta)=\frac{GV_0}{2} & \sum\limits_{m_x}\sum\limits_{m_l}[P_i(\Delta+\delta+(m_l+m_x)A_{||}) \\
& -P_i(\Delta-\delta+(m_l+m_x)A_{||})],
\end{split}
\label{Si}
\end{equation}
where $G$ is the lock-in gain factor, $V_0$ is the off-resonant detected voltage, and $P_i$ is either the reflected or transmitted cavity power. The expression is summed over the $^{14}$N nuclear spin quantum number \mbox{$m_l$ = $\{$-1,0,1$\}$}, and the three frequencies $m_x$ separated by $A_{||}$ in order to account for the simultaneous drive of all three hyperfine transitions.

Using \mbox{Eq. (\ref{Si})}, $S^{P_t}_{LI}$ is plotted as a solid line in \mbox{Fig. \ref{map}(a)}, taking \mbox{$P_{in}$ = 0.4 W}, \mbox{$\Omega$ = 0.3 MHz}, a pure dephasing rate of \mbox{$\gamma^*_2$ = 1/3 MHz}, a longitudinal relaxation rate of \mbox{$\gamma_1$ = 0.182 kHz}, and level decay rates $k_{ab}$ extracted from \cite{Robledo2011a}.
We also plot the measured ADMR spectrum as a dashed line in \mbox{Fig. \ref{map}(a)}. The ADMR spectrum was recorded with the same $P_{in}$ and $\Omega$ as the simulated spectrum. 
The match between the simulated and measured traces is very good, with just a small mismatch due to the uncertainty in the estimation of the parameters $\epsilon$, $\gamma^*_2$, and $\alpha^0_{abs}$ in the simulation.

\begin{figure*}
	\includegraphics[scale = 0.3]{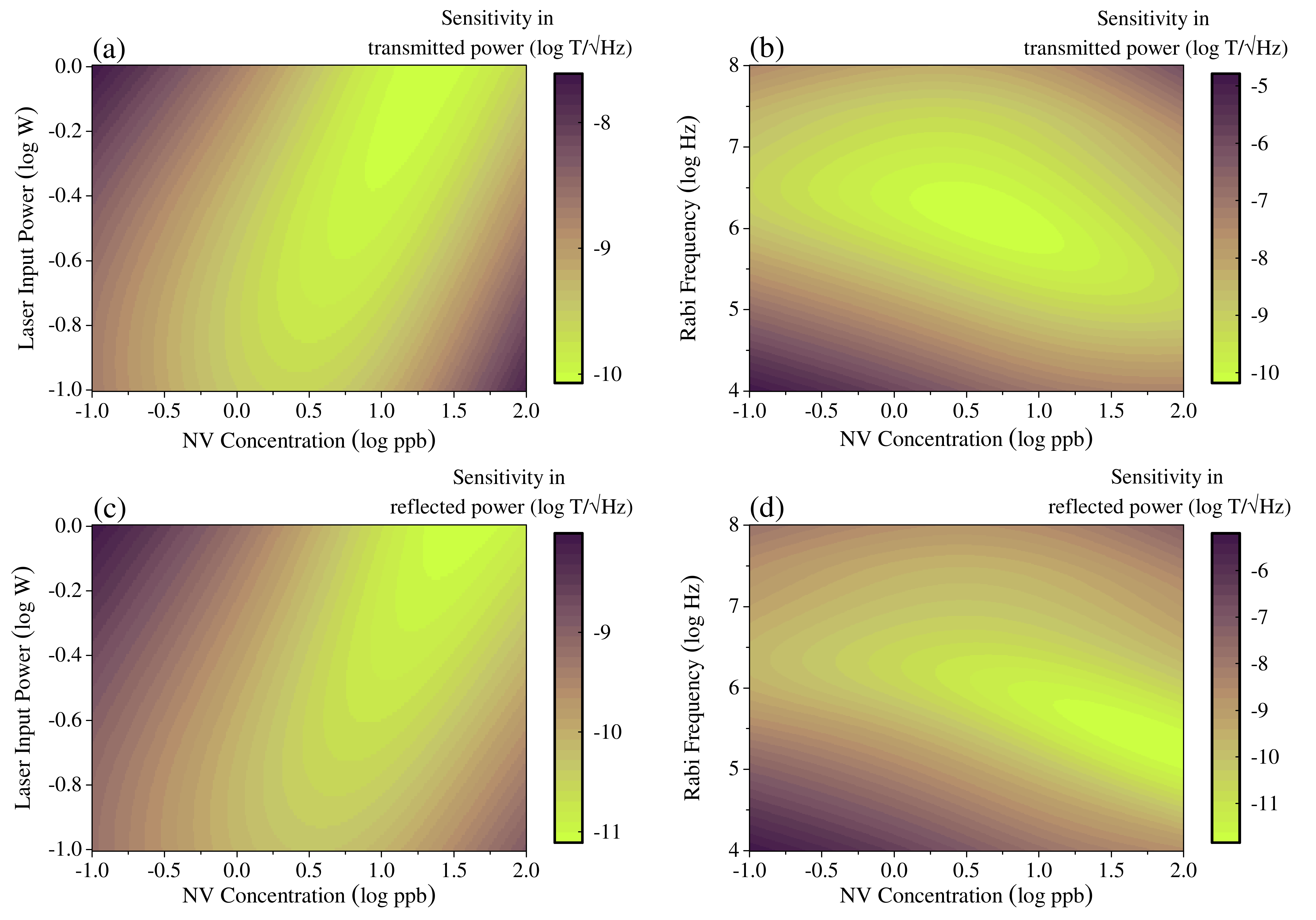}
	\caption{Simulated plots of the shot-noise-limited sensitivity as a function of [NV$^-$], $P_{in}$, and $\Omega$ using the following parameters
\mbox{$R_1=R_2e^{-\alpha}$}, \mbox{$R_2$ = 99.9 $\%$}, \mbox{$\alpha^0_{abs}=\alpha^{\text{NV}}_{abs}$}, \mbox{$\alpha_r$ = 0.006}, \mbox{$l$ = 2$\times$1.3 mm}, \mbox{$\epsilon$ = 75 kHz/W}, \mbox{$\gamma^*_2$ = 1/3 MHz}, \mbox{$\gamma_1$ = 0.182 kHz}.
(a) and (b) are calculated from transmission through the cavity for \mbox{$\Omega$ = 0.5 MHz} and \mbox{$P_{in}$ = 0.5 W}, respectively. (c) and (d) are calculated from reflection of the cavity for \mbox{$\Omega$ = 0.5 MHz} and \mbox{$P_{in}$ = 0.5 W}, respectively.}
	\label{Density}
\end{figure*}

\subsection{Sensitivity}
To optimize the magnetic field sensitivity, we measure the dependence of $\frac{d}{d_w}(S^{P_t}_{LI})$ of three-frequency excitation spectra on the pump power and Rabi frequency, $P_{in}$ and $\Omega$, at \mbox{$\Delta=0$}.
The results of these measurements are presented in \mbox{Fig. \ref{map}(b)}. The maximum slope is achieved at \mbox{$P_{in}$ = 0.4 W} and \mbox{$\Omega\sim$ 0.3 MHz}, where the optical excitation rate by virtue of the cavity enhancement overcomes the MW power-induced broadening, allowing for a narrowing regime to be reached \cite{Jensen2013}. The simulated slopes are presented in \mbox{Fig. \ref{map}(c)} and obtained using the same parameters as in \mbox{Fig. \ref{map}(a)}. We observe a very good agreement with respect to the overall trend, the slope magnitude, and the location of the slope maximum.

For deducing the sensitivity of the magnetometer, we independently measured four time traces of the lock-in signal for \mbox{$P_{in}$ = 0.4 W}. The first trace was measured in the optimal magnetically sensitive configuration, with the MW drive on resonance with a spin transition (\mbox{$\Delta=0$}) corresponding to the purple dot in \mbox{Fig. \ref{ODMR}(b)}. The second trace was measured in the magnetically insensitive configuration, with the MW drive frequency far-detuned from any spin resonance (\mbox{$\Delta \rightarrow \infty$}). The third trace was measured by blocking the reference detector which monitored the laser output. The last trace was measured with all detectors blocked, which shows the sum of electronic noise from the lock-in detector and photodetectors.
The Fourier transforms of these time traces with a frequency resolution of \mbox{0.24 Hz} are presented in \mbox{Fig. \ref{Pulsed}(a)} where the y-axis is displayed in units of sensitivity.
It shows a \mbox{125 Hz} bandwidth and a \mbox{12 dB/octave} roll-off that is generated by the low-pass filter of the lock-in detector. The choice of this bandwidth is a consequence of the low ADMR contrast (\mbox{$C_{\text{ADMR}} \sim 10^{-6}$}) measured through the cavity transmission.
When the MW drive is off resonance, a noise floor of \mbox{$\sim$ 100 nT/$\sqrt{\textrm{Hz}}$} is achieved.
The increased noise floor when we blocked the reference detector and only monitored the transmission through the cavity shows the impact of substantial technical noise at the \mbox{35 kHz} modulation frequency.
Next, we calculated the Allan deviation of both magnetically insensitive and magnetically sensitive traces, which allows us to investigate the intrinsic noise in the system. The results are presented in \mbox{Fig \ref{Pulsed}(b)}. The drop of the Allan deviation with a slope of -1/2 in both traces is a signature of white noise. For the magnetically sensitive measurements, the white noise reaches a minimum at \mbox{$\sim$ 3.3 s}. The increase of the Allan deviation at higher averaging time is a sign of thermal or mechanical drift in the system.

\section{Outlook}
To better understand the context and magnitude of the measured sensitivity, we estimate the shot-noise-limited sensitivity for a single-peak ADMR as a function of [NV$^-$], $P_{in}$, and $\Omega$. Using the same physical dimensions as in our setup (cavity length and diamond thickness), we assume a diamond host where \mbox{$\alpha^0_{abs}=\alpha^{\text{NV}}_{abs}$} for any [NV$^-$]. In addition, we consider that the reflectivity of the incoupling mirror is such that \mbox{$R_1=R_2e^{-\alpha}$} when \mbox{$\Omega=0$} for a given optical input power, ensuring that the cavity is impedance-matched. The intra-cavity power is thereby always maximized and there is no cavity reflection when no MW field is applied. The shot-noise-limited sensitivity was estimated from the ratio of the shot-noise level to max[$\frac{d}{d_w}S$]. The results of this calculation are presented in \mbox{Fig. \ref{Density}} for both transmitted (a,b) and reflected (c,d) powers. We have fixed \mbox{$\Omega=0.5$ MHz} for (a,c) and \mbox{$P_{in}=0.5$ W} for (b,d). By monitoring the transmitted power $P_t$ and optimizing [NV$^-$], $P_{in}$, and $\Omega$, a shot-noise-limited sensitivity in the \mbox{sub-100-pT/$\sqrt{\textrm{Hz}}$} range can be expected. In comparison, by monitoring the reflected power $P_r$, a sensitivity in the \mbox{pT/$\sqrt{\textrm{Hz}}$} range is projected. As the cavity is impedance-matched, applying no MW field results in \mbox{$P_r(\Omega=0)\sim0$}. However, applying $\Omega$ on resonance with a spin transition reduces the loss in the cavity and pushes the cavity into the over-coupled regime. For the case presented in \mbox{Fig. \ref{Density}(d)} with a fixed input power \mbox{$P_{in}=0.5$ W}, the optimal sensitivity of \mbox{$\sim$ 1 pT/$\sqrt{\textrm{Hz}}$} is obtained for \mbox{$\Omega = 0.21$ MHz} and \mbox{[NV$^-$] $\sim$ 70.8 ppb}. At these settings, the cavity finesse is 13.7, the intra-cavity power reaches \mbox{$P_{cav}=5.35$ W} and the maximum reflected power \mbox{$P_r(\Delta=0)$ = 0.15 $\mu$W }. The total reflected power of such an over-coupled cavity contributes to the ADMR signal. 

\section{Conclusion}
In this article, we report on magnetic field sensing using an ensemble of NV centers based on the variation of a cavity's transmitted pump power due to electron-spin absorption. Frequency-modulated ADMR spectra were measured, which was used to measure the local magnetic noise spectral density with a noise floor of \mbox{100 nT/$\sqrt{\textrm{Hz}}$} spanning a bandwidth up to \mbox{125 Hz}. Our simulations show that a photon shot-noise-limited sensitivity of \mbox{$\sim$ 1 pT/$\sqrt{\textrm{Hz}}$} can be achieved when measuring a cavity's reflected power near the impedance-matched point and using a diamond with an optimized NV density. Cavity-based ADMR is an alternative to its ODMR counterpart, and has advantageous in terms of both detection contrast and device application.
With the appropriate cavity design and sample optimization, it is anticipated that the work and technique presented here will provide a solid foundation for NV-based magnetometers.

\section{Acknowledgments}
We would like to thank Kristian Hagsted Rasmussen for help with the diamond sample preparation. We are also grateful to Jonas Schou Neergaard-Nielsen for fruitful discussions. This work was supported by the Danish Innovation Foundation through the EXMAD project and the Qubiz center, as well as the Danish Research Council through the Sapere Aude project (DIMS).

\bibliography{library}

\begin{thebibliography}{32}%
\makeatletter
\providecommand \@ifxundefined [1]{%
 \@ifx{#1\undefined}
}%
\providecommand \@ifnum [1]{%
 \ifnum #1\expandafter \@firstoftwo
 \else \expandafter \@secondoftwo
 \fi
}%
\providecommand \@ifx [1]{%
 \ifx #1\expandafter \@firstoftwo
 \else \expandafter \@secondoftwo
 \fi
}%
\providecommand \natexlab [1]{#1}%
\providecommand \enquote  [1]{``#1''}%
\providecommand \bibnamefont  [1]{#1}%
\providecommand \bibfnamefont [1]{#1}%
\providecommand \citenamefont [1]{#1}%
\providecommand \href@noop [0]{\@secondoftwo}%
\providecommand \href [0]{\begingroup \@sanitize@url \@href}%
\providecommand \@href[1]{\@@startlink{#1}\@@href}%
\providecommand \@@href[1]{\endgroup#1\@@endlink}%
\providecommand \@sanitize@url [0]{\catcode `\\12\catcode `\$12\catcode
  `\&12\catcode `\#12\catcode `\^12\catcode `\_12\catcode `\%12\relax}%
\providecommand \@@startlink[1]{}%
\providecommand \@@endlink[0]{}%
\providecommand \url  [0]{\begingroup\@sanitize@url \@url }%
\providecommand \@url [1]{\endgroup\@href {#1}{\urlprefix }}%
\providecommand \urlprefix  [0]{URL }%
\providecommand \Eprint [0]{\href }%
\providecommand \doibase [0]{http://dx.doi.org/}%
\providecommand \selectlanguage [0]{\@gobble}%
\providecommand \bibinfo  [0]{\@secondoftwo}%
\providecommand \bibfield  [0]{\@secondoftwo}%
\providecommand \translation [1]{[#1]}%
\providecommand \BibitemOpen [0]{}%
\providecommand \bibitemStop [0]{}%
\providecommand \bibitemNoStop [0]{.\EOS\space}%
\providecommand \EOS [0]{\spacefactor3000\relax}%
\providecommand \BibitemShut  [1]{\csname bibitem#1\endcsname}%
\let\auto@bib@innerbib\@empty
\bibitem [{\citenamefont {Balasubramanian}\ \emph {et~al.}(2008)\citenamefont
  {Balasubramanian}, \citenamefont {Chan}, \citenamefont {Kolesov},
  \citenamefont {Al-Hmoud}, \citenamefont {Tisler}, \citenamefont {Shin},
  \citenamefont {Kim}, \citenamefont {Wojcik}, \citenamefont {Hemmer},
  \citenamefont {Krueger}, \citenamefont {Hanke}, \citenamefont
  {Leitenstorfer}, \citenamefont {Bratschitsch}, \citenamefont {Jelezko},\ and\
  \citenamefont {Wrachtrup}}]{Balasubramanian2008}%
  \BibitemOpen
  \bibfield  {author} {\bibinfo {author} {\bibfnamefont {G.}~\bibnamefont
  {Balasubramanian}}, \bibinfo {author} {\bibfnamefont {I.~Y.}\ \bibnamefont
  {Chan}}, \bibinfo {author} {\bibfnamefont {R.}~\bibnamefont {Kolesov}},
  \bibinfo {author} {\bibfnamefont {M.}~\bibnamefont {Al-Hmoud}}, \bibinfo
  {author} {\bibfnamefont {J.}~\bibnamefont {Tisler}}, \bibinfo {author}
  {\bibfnamefont {C.}~\bibnamefont {Shin}}, \bibinfo {author} {\bibfnamefont
  {C.}~\bibnamefont {Kim}}, \bibinfo {author} {\bibfnamefont {A.}~\bibnamefont
  {Wojcik}}, \bibinfo {author} {\bibfnamefont {P.~R.}\ \bibnamefont {Hemmer}},
  \bibinfo {author} {\bibfnamefont {A.}~\bibnamefont {Krueger}}, \bibinfo
  {author} {\bibfnamefont {T.}~\bibnamefont {Hanke}}, \bibinfo {author}
  {\bibfnamefont {A.}~\bibnamefont {Leitenstorfer}}, \bibinfo {author}
  {\bibfnamefont {R.}~\bibnamefont {Bratschitsch}}, \bibinfo {author}
  {\bibfnamefont {F.}~\bibnamefont {Jelezko}}, \ and\ \bibinfo {author}
  {\bibfnamefont {J.}~\bibnamefont {Wrachtrup}},\ }\href {\doibase
  10.1038/nature07278} {\bibfield  {journal} {\bibinfo  {journal} {Nature}\
  }\textbf {\bibinfo {volume} {455}},\ \bibinfo {pages} {648} (\bibinfo {year}
  {2008})}\BibitemShut {NoStop}%
\bibitem [{\citenamefont {Rondin}\ \emph {et~al.}(2014)\citenamefont {Rondin},
  \citenamefont {Tetienne}, \citenamefont {Hingant}, \citenamefont {Roch},
  \citenamefont {Maletinsky},\ and\ \citenamefont {Jacques}}]{Rondin2014}%
  \BibitemOpen
  \bibfield  {author} {\bibinfo {author} {\bibfnamefont {L.}~\bibnamefont
  {Rondin}}, \bibinfo {author} {\bibfnamefont {J.~P.}\ \bibnamefont
  {Tetienne}}, \bibinfo {author} {\bibfnamefont {T.}~\bibnamefont {Hingant}},
  \bibinfo {author} {\bibfnamefont {J.~F.}\ \bibnamefont {Roch}}, \bibinfo
  {author} {\bibfnamefont {P.}~\bibnamefont {Maletinsky}}, \ and\ \bibinfo
  {author} {\bibfnamefont {V.}~\bibnamefont {Jacques}},\ }\href {\doibase
  10.1088/0034-4885/77/5/056503} {\bibfield  {journal} {\bibinfo  {journal}
  {Reports on Progress in Physics}\ }\textbf {\bibinfo {volume} {77}},\
  \bibinfo {pages} {56503} (\bibinfo {year} {2014})}\BibitemShut {NoStop}%
\bibitem [{\citenamefont {Schirhagl}\ \emph {et~al.}(2014)\citenamefont
  {Schirhagl}, \citenamefont {Chang}, \citenamefont {Loretz},\ and\
  \citenamefont {Degen}}]{Schirhagl2014}%
  \BibitemOpen
  \bibfield  {author} {\bibinfo {author} {\bibfnamefont {R.}~\bibnamefont
  {Schirhagl}}, \bibinfo {author} {\bibfnamefont {K.}~\bibnamefont {Chang}},
  \bibinfo {author} {\bibfnamefont {M.}~\bibnamefont {Loretz}}, \ and\ \bibinfo
  {author} {\bibfnamefont {C.~L.}\ \bibnamefont {Degen}},\ }\href {\doibase
  10.1146/annurev-physchem-040513-103659} {\bibfield  {journal} {\bibinfo
  {journal} {Annual Review of Physical Chemistry}\ }\textbf {\bibinfo {volume}
  {65}},\ \bibinfo {pages} {83} (\bibinfo {year} {2014})}\BibitemShut {NoStop}%
\bibitem [{\citenamefont {Dolde}\ \emph {et~al.}(2011)\citenamefont {Dolde},
  \citenamefont {Fedder}, \citenamefont {Doherty}, \citenamefont
  {N{\"{o}}bauer}, \citenamefont {Rempp}, \citenamefont {Balasubramanian},
  \citenamefont {Wolf}, \citenamefont {Reinhard}, \citenamefont {Hollenberg},
  \citenamefont {Jelezko},\ and\ \citenamefont {Wrachtrup}}]{Dolde2011}%
  \BibitemOpen
  \bibfield  {author} {\bibinfo {author} {\bibfnamefont {F.}~\bibnamefont
  {Dolde}}, \bibinfo {author} {\bibfnamefont {H.}~\bibnamefont {Fedder}},
  \bibinfo {author} {\bibfnamefont {M.~W.}\ \bibnamefont {Doherty}}, \bibinfo
  {author} {\bibfnamefont {T.}~\bibnamefont {N{\"{o}}bauer}}, \bibinfo {author}
  {\bibfnamefont {F.}~\bibnamefont {Rempp}}, \bibinfo {author} {\bibfnamefont
  {G.}~\bibnamefont {Balasubramanian}}, \bibinfo {author} {\bibfnamefont
  {T.}~\bibnamefont {Wolf}}, \bibinfo {author} {\bibfnamefont {F.}~\bibnamefont
  {Reinhard}}, \bibinfo {author} {\bibfnamefont {L.~C.~L.}\ \bibnamefont
  {Hollenberg}}, \bibinfo {author} {\bibfnamefont {F.}~\bibnamefont {Jelezko}},
  \ and\ \bibinfo {author} {\bibfnamefont {J.}~\bibnamefont {Wrachtrup}},\
  }\href {\doibase 10.1038/nphys1969} {\bibfield  {journal} {\bibinfo
  {journal} {Nature Physics}\ }\textbf {\bibinfo {volume} {7}},\ \bibinfo
  {pages} {459} (\bibinfo {year} {2011})}\BibitemShut {NoStop}%
\bibitem [{\citenamefont {Kucsko}\ \emph {et~al.}(2013)\citenamefont {Kucsko},
  \citenamefont {Maurer}, \citenamefont {Yao}, \citenamefont {Kubo},
  \citenamefont {Noh}, \citenamefont {Lo}, \citenamefont {Park},\ and\
  \citenamefont {Lukin}}]{Kucsko2013}%
  \BibitemOpen
  \bibfield  {author} {\bibinfo {author} {\bibfnamefont {G.}~\bibnamefont
  {Kucsko}}, \bibinfo {author} {\bibfnamefont {P.~C.}\ \bibnamefont {Maurer}},
  \bibinfo {author} {\bibfnamefont {N.~Y.}\ \bibnamefont {Yao}}, \bibinfo
  {author} {\bibfnamefont {M.}~\bibnamefont {Kubo}}, \bibinfo {author}
  {\bibfnamefont {H.~J.}\ \bibnamefont {Noh}}, \bibinfo {author} {\bibfnamefont
  {P.~K.}\ \bibnamefont {Lo}}, \bibinfo {author} {\bibfnamefont
  {H.}~\bibnamefont {Park}}, \ and\ \bibinfo {author} {\bibfnamefont {M.~D.}\
  \bibnamefont {Lukin}},\ }\href {\doibase 10.1038/nature12373} {\bibfield
  {journal} {\bibinfo  {journal} {Nature}\ }\textbf {\bibinfo {volume} {500}},\
  \bibinfo {pages} {54} (\bibinfo {year} {2013})}\BibitemShut {NoStop}%
\bibitem [{\citenamefont {Neumann}\ \emph {et~al.}(2013)\citenamefont
  {Neumann}, \citenamefont {Jakobi}, \citenamefont {Dolde}, \citenamefont
  {Burk}, \citenamefont {Reuter}, \citenamefont {Waldherr}, \citenamefont
  {Honert}, \citenamefont {Wolf}, \citenamefont {Brunner}, \citenamefont
  {Shim}, \citenamefont {Suter}, \citenamefont {Sumiya}, \citenamefont
  {Isoya},\ and\ \citenamefont {Wrachtrup}}]{Neumann2013}%
  \BibitemOpen
  \bibfield  {author} {\bibinfo {author} {\bibfnamefont {P.}~\bibnamefont
  {Neumann}}, \bibinfo {author} {\bibfnamefont {I.}~\bibnamefont {Jakobi}},
  \bibinfo {author} {\bibfnamefont {F.}~\bibnamefont {Dolde}}, \bibinfo
  {author} {\bibfnamefont {C.}~\bibnamefont {Burk}}, \bibinfo {author}
  {\bibfnamefont {R.}~\bibnamefont {Reuter}}, \bibinfo {author} {\bibfnamefont
  {G.}~\bibnamefont {Waldherr}}, \bibinfo {author} {\bibfnamefont
  {J.}~\bibnamefont {Honert}}, \bibinfo {author} {\bibfnamefont
  {T.}~\bibnamefont {Wolf}}, \bibinfo {author} {\bibfnamefont {A.}~\bibnamefont
  {Brunner}}, \bibinfo {author} {\bibfnamefont {J.~H.}\ \bibnamefont {Shim}},
  \bibinfo {author} {\bibfnamefont {D.}~\bibnamefont {Suter}}, \bibinfo
  {author} {\bibfnamefont {H.}~\bibnamefont {Sumiya}}, \bibinfo {author}
  {\bibfnamefont {J.}~\bibnamefont {Isoya}}, \ and\ \bibinfo {author}
  {\bibfnamefont {J.}~\bibnamefont {Wrachtrup}},\ }\href
  {http://pubs.acs.org/doi/pdf/10.1021/nl401216y} {\bibfield  {journal}
  {\bibinfo  {journal} {Nano Letters}\ }\textbf {\bibinfo {volume} {13}},\
  \bibinfo {pages} {2738} (\bibinfo {year} {2013})}\BibitemShut {NoStop}%
\bibitem [{\citenamefont {Hall}\ \emph {et~al.}(2012)\citenamefont {Hall},
  \citenamefont {Beart}, \citenamefont {Thomas}, \citenamefont {Simpson},
  \citenamefont {McGuinness}, \citenamefont {Cole}, \citenamefont {Manton},
  \citenamefont {Scholten}, \citenamefont {Jelezko}, \citenamefont {Wrachtrup},
  \citenamefont {Petrou},\ and\ \citenamefont {Hollenberg}}]{Hall2012}%
  \BibitemOpen
  \bibfield  {author} {\bibinfo {author} {\bibfnamefont {L.~T.}\ \bibnamefont
  {Hall}}, \bibinfo {author} {\bibfnamefont {G.~C.~G.}\ \bibnamefont {Beart}},
  \bibinfo {author} {\bibfnamefont {E.~A.}\ \bibnamefont {Thomas}}, \bibinfo
  {author} {\bibfnamefont {D.~A.}\ \bibnamefont {Simpson}}, \bibinfo {author}
  {\bibfnamefont {L.~P.}\ \bibnamefont {McGuinness}}, \bibinfo {author}
  {\bibfnamefont {J.~H.}\ \bibnamefont {Cole}}, \bibinfo {author}
  {\bibfnamefont {J.~H.}\ \bibnamefont {Manton}}, \bibinfo {author}
  {\bibfnamefont {R.~E.}\ \bibnamefont {Scholten}}, \bibinfo {author}
  {\bibfnamefont {F.}~\bibnamefont {Jelezko}}, \bibinfo {author} {\bibfnamefont
  {J.}~\bibnamefont {Wrachtrup}}, \bibinfo {author} {\bibfnamefont
  {S.}~\bibnamefont {Petrou}}, \ and\ \bibinfo {author} {\bibfnamefont
  {L.~C.~L.}\ \bibnamefont {Hollenberg}},\ }\href {\doibase 10.1038/srep00401}
  {\bibfield  {journal} {\bibinfo  {journal} {Scientific reports}\ }\textbf
  {\bibinfo {volume} {2}},\ \bibinfo {pages} {1} (\bibinfo {year}
  {2012})}\BibitemShut {NoStop}%
\bibitem [{\citenamefont {Barry}\ \emph {et~al.}(2016)\citenamefont {Barry},
  \citenamefont {Turner}, \citenamefont {Schloss}, \citenamefont {Glenn},
  \citenamefont {Song}, \citenamefont {Lukin}, \citenamefont {Park},\ and\
  \citenamefont {Walsworth}}]{Barry2016}%
  \BibitemOpen
  \bibfield  {author} {\bibinfo {author} {\bibfnamefont {J.~F.}\ \bibnamefont
  {Barry}}, \bibinfo {author} {\bibfnamefont {M.~J.}\ \bibnamefont {Turner}},
  \bibinfo {author} {\bibfnamefont {J.~M.}\ \bibnamefont {Schloss}}, \bibinfo
  {author} {\bibfnamefont {D.~R.}\ \bibnamefont {Glenn}}, \bibinfo {author}
  {\bibfnamefont {Y.}~\bibnamefont {Song}}, \bibinfo {author} {\bibfnamefont
  {M.~D.}\ \bibnamefont {Lukin}}, \bibinfo {author} {\bibfnamefont
  {H.}~\bibnamefont {Park}}, \ and\ \bibinfo {author} {\bibfnamefont {R.~L.}\
  \bibnamefont {Walsworth}},\ }\href {\doibase 10.1073/pnas.1601513113}
  {\bibfield  {journal} {\bibinfo  {journal} {Proceedings of the National
  Academy of Sciences}\ }\textbf {\bibinfo {volume} {113}},\ \bibinfo {pages}
  {14133} (\bibinfo {year} {2016})}\BibitemShut {NoStop}%
\bibitem [{\citenamefont {Steinert}\ \emph {et~al.}(2013)\citenamefont
  {Steinert}, \citenamefont {Ziem}, \citenamefont {Hall}, \citenamefont
  {Zappe}, \citenamefont {Schweikert}, \citenamefont {G{\"{o}}tz},
  \citenamefont {Aird}, \citenamefont {Balasubramanian}, \citenamefont
  {Hollenberg},\ and\ \citenamefont {Wrachtrup}}]{Steinert2013}%
  \BibitemOpen
  \bibfield  {author} {\bibinfo {author} {\bibfnamefont {S.}~\bibnamefont
  {Steinert}}, \bibinfo {author} {\bibfnamefont {F.}~\bibnamefont {Ziem}},
  \bibinfo {author} {\bibfnamefont {L.~T.}\ \bibnamefont {Hall}}, \bibinfo
  {author} {\bibfnamefont {A.}~\bibnamefont {Zappe}}, \bibinfo {author}
  {\bibfnamefont {M.}~\bibnamefont {Schweikert}}, \bibinfo {author}
  {\bibfnamefont {N.}~\bibnamefont {G{\"{o}}tz}}, \bibinfo {author}
  {\bibfnamefont {A.}~\bibnamefont {Aird}}, \bibinfo {author} {\bibfnamefont
  {G.}~\bibnamefont {Balasubramanian}}, \bibinfo {author} {\bibfnamefont
  {L.}~\bibnamefont {Hollenberg}}, \ and\ \bibinfo {author} {\bibfnamefont
  {J.}~\bibnamefont {Wrachtrup}},\ }\href {\doibase 10.1038/ncomms2588}
  {\bibfield  {journal} {\bibinfo  {journal} {Nature Communications}\ }\textbf
  {\bibinfo {volume} {4}},\ \bibinfo {pages} {1607} (\bibinfo {year}
  {2013})}\BibitemShut {NoStop}%
\bibitem [{\citenamefont {Glenn}\ \emph {et~al.}(2015)\citenamefont {Glenn},
  \citenamefont {Lee}, \citenamefont {Park}, \citenamefont {Weissleder},
  \citenamefont {Yacoby}, \citenamefont {Lukin}, \citenamefont {Lee},
  \citenamefont {Walsworth},\ and\ \citenamefont {Connolly}}]{Glenn2015}%
  \BibitemOpen
  \bibfield  {author} {\bibinfo {author} {\bibfnamefont {D.~R.}\ \bibnamefont
  {Glenn}}, \bibinfo {author} {\bibfnamefont {K.}~\bibnamefont {Lee}}, \bibinfo
  {author} {\bibfnamefont {H.}~\bibnamefont {Park}}, \bibinfo {author}
  {\bibfnamefont {R.}~\bibnamefont {Weissleder}}, \bibinfo {author}
  {\bibfnamefont {A.}~\bibnamefont {Yacoby}}, \bibinfo {author} {\bibfnamefont
  {M.~D.}\ \bibnamefont {Lukin}}, \bibinfo {author} {\bibfnamefont
  {H.}~\bibnamefont {Lee}}, \bibinfo {author} {\bibfnamefont {R.~L.}\
  \bibnamefont {Walsworth}}, \ and\ \bibinfo {author} {\bibfnamefont {C.~B.}\
  \bibnamefont {Connolly}},\ }\href {\doibase 10.1038/nmeth.3449} {\bibfield
  {journal} {\bibinfo  {journal} {Nature Methods}\ }\textbf {\bibinfo {volume}
  {12}},\ \bibinfo {pages} {736} (\bibinfo {year} {2015})}\BibitemShut
  {NoStop}%
\bibitem [{\citenamefont {Lovchinsky}\ \emph {et~al.}(2016)\citenamefont
  {Lovchinsky}, \citenamefont {Sushkov}, \citenamefont {Urbach}, \citenamefont
  {de~Leon}, \citenamefont {Choi}, \citenamefont {Greve}, \citenamefont
  {Evans}, \citenamefont {Gertner}, \citenamefont {Bersin}, \citenamefont
  {M{\"{u}}ller}, \citenamefont {McGuinness}, \citenamefont {Jelezko},
  \citenamefont {Walsworth}, \citenamefont {Park},\ and\ \citenamefont
  {Lukin}}]{Lovchinsky2016}%
  \BibitemOpen
  \bibfield  {author} {\bibinfo {author} {\bibfnamefont {I.}~\bibnamefont
  {Lovchinsky}}, \bibinfo {author} {\bibfnamefont {A.~O.}\ \bibnamefont
  {Sushkov}}, \bibinfo {author} {\bibfnamefont {E.}~\bibnamefont {Urbach}},
  \bibinfo {author} {\bibfnamefont {N.~P.}\ \bibnamefont {de~Leon}}, \bibinfo
  {author} {\bibfnamefont {S.}~\bibnamefont {Choi}}, \bibinfo {author}
  {\bibfnamefont {K.~D.}\ \bibnamefont {Greve}}, \bibinfo {author}
  {\bibfnamefont {R.}~\bibnamefont {Evans}}, \bibinfo {author} {\bibfnamefont
  {R.}~\bibnamefont {Gertner}}, \bibinfo {author} {\bibfnamefont
  {E.}~\bibnamefont {Bersin}}, \bibinfo {author} {\bibfnamefont
  {C.}~\bibnamefont {M{\"{u}}ller}}, \bibinfo {author} {\bibfnamefont
  {L.}~\bibnamefont {McGuinness}}, \bibinfo {author} {\bibfnamefont
  {F.}~\bibnamefont {Jelezko}}, \bibinfo {author} {\bibfnamefont {R.~L.}\
  \bibnamefont {Walsworth}}, \bibinfo {author} {\bibfnamefont {H.}~\bibnamefont
  {Park}}, \ and\ \bibinfo {author} {\bibfnamefont {M.~D.}\ \bibnamefont
  {Lukin}},\ }\href@noop {} {\bibfield  {journal} {\bibinfo  {journal}
  {Science}\ }\textbf {\bibinfo {volume} {351}},\ \bibinfo {pages} {836}
  (\bibinfo {year} {2016})}\BibitemShut {NoStop}%
\bibitem [{\citenamefont {Fu}\ \emph {et~al.}(2014)\citenamefont {Fu},
  \citenamefont {Weiss}, \citenamefont {Lima}, \citenamefont {Harrison},
  \citenamefont {Bai}, \citenamefont {Desch}, \citenamefont {Ebel},
  \citenamefont {Suavet}, \citenamefont {Wang}, \citenamefont {Glenn},
  \citenamefont {Sage}, \citenamefont {Kasama}, \citenamefont {Walsworth},\
  and\ \citenamefont {Kuan}}]{Fu2014}%
  \BibitemOpen
  \bibfield  {author} {\bibinfo {author} {\bibfnamefont {R.~R.}\ \bibnamefont
  {Fu}}, \bibinfo {author} {\bibfnamefont {B.~P.}\ \bibnamefont {Weiss}},
  \bibinfo {author} {\bibfnamefont {E.~A.}\ \bibnamefont {Lima}}, \bibinfo
  {author} {\bibfnamefont {R.~J.}\ \bibnamefont {Harrison}}, \bibinfo {author}
  {\bibfnamefont {X.-N.}\ \bibnamefont {Bai}}, \bibinfo {author} {\bibfnamefont
  {S.~J.}\ \bibnamefont {Desch}}, \bibinfo {author} {\bibfnamefont {D.~S.}\
  \bibnamefont {Ebel}}, \bibinfo {author} {\bibfnamefont {C.}~\bibnamefont
  {Suavet}}, \bibinfo {author} {\bibfnamefont {H.}~\bibnamefont {Wang}},
  \bibinfo {author} {\bibfnamefont {D.}~\bibnamefont {Glenn}}, \bibinfo
  {author} {\bibfnamefont {D.~L.}\ \bibnamefont {Sage}}, \bibinfo {author}
  {\bibfnamefont {T.}~\bibnamefont {Kasama}}, \bibinfo {author} {\bibfnamefont
  {R.~L.}\ \bibnamefont {Walsworth}}, \ and\ \bibinfo {author} {\bibfnamefont
  {A.~T.}\ \bibnamefont {Kuan}},\ }\href@noop {} {\bibfield  {journal}
  {\bibinfo  {journal} {Science}\ }\textbf {\bibinfo {volume} {346}},\ \bibinfo
  {pages} {1089} (\bibinfo {year} {2014})}\BibitemShut {NoStop}%
\bibitem [{\citenamefont {Kolkowitz}\ \emph {et~al.}(2015)\citenamefont
  {Kolkowitz}, \citenamefont {Safira}, \citenamefont {High}, \citenamefont
  {Devlin}, \citenamefont {Choi}, \citenamefont {Unterreithmeier},
  \citenamefont {Patterson}, \citenamefont {Zibrov}, \citenamefont
  {Manucharyan}, \citenamefont {Park},\ and\ \citenamefont
  {Lukin}}]{Kolkowitz2015}%
  \BibitemOpen
  \bibfield  {author} {\bibinfo {author} {\bibfnamefont {S.}~\bibnamefont
  {Kolkowitz}}, \bibinfo {author} {\bibfnamefont {A.}~\bibnamefont {Safira}},
  \bibinfo {author} {\bibfnamefont {A.~A.}\ \bibnamefont {High}}, \bibinfo
  {author} {\bibfnamefont {R.~C.}\ \bibnamefont {Devlin}}, \bibinfo {author}
  {\bibfnamefont {S.}~\bibnamefont {Choi}}, \bibinfo {author} {\bibfnamefont
  {Q.~P.}\ \bibnamefont {Unterreithmeier}}, \bibinfo {author} {\bibfnamefont
  {D.}~\bibnamefont {Patterson}}, \bibinfo {author} {\bibfnamefont {A.~S.}\
  \bibnamefont {Zibrov}}, \bibinfo {author} {\bibfnamefont {V.~E.}\
  \bibnamefont {Manucharyan}}, \bibinfo {author} {\bibfnamefont
  {H.}~\bibnamefont {Park}}, \ and\ \bibinfo {author} {\bibfnamefont {M.~D.}\
  \bibnamefont {Lukin}},\ }\href {\doibase 10.1126/science.aaa4298} {\bibfield
  {journal} {\bibinfo  {journal} {Science}\ }\textbf {\bibinfo {volume}
  {347}},\ \bibinfo {pages} {1129} (\bibinfo {year} {2015})}\BibitemShut
  {NoStop}%
\bibitem [{\citenamefont {Jakobi}\ \emph {et~al.}(2016)\citenamefont {Jakobi},
  \citenamefont {Neumann}, \citenamefont {Wang}, \citenamefont {Dasari},
  \citenamefont {Hallak}, \citenamefont {Bashir}, \citenamefont {Markham},
  \citenamefont {Edmonds}, \citenamefont {Twitchen},\ and\ \citenamefont
  {Wrachtrup}}]{Jakobi2016}%
  \BibitemOpen
  \bibfield  {author} {\bibinfo {author} {\bibfnamefont {I.}~\bibnamefont
  {Jakobi}}, \bibinfo {author} {\bibfnamefont {P.}~\bibnamefont {Neumann}},
  \bibinfo {author} {\bibfnamefont {Y.}~\bibnamefont {Wang}}, \bibinfo {author}
  {\bibfnamefont {D.}~\bibnamefont {Dasari}}, \bibinfo {author} {\bibfnamefont
  {F.~E.}\ \bibnamefont {Hallak}}, \bibinfo {author} {\bibfnamefont {M.~A.}\
  \bibnamefont {Bashir}}, \bibinfo {author} {\bibfnamefont {M.}~\bibnamefont
  {Markham}}, \bibinfo {author} {\bibfnamefont {A.}~\bibnamefont {Edmonds}},
  \bibinfo {author} {\bibfnamefont {D.}~\bibnamefont {Twitchen}}, \ and\
  \bibinfo {author} {\bibfnamefont {J.}~\bibnamefont {Wrachtrup}},\ }\href
  {\doibase 10.1038/nnano.2016.163} {\bibfield  {journal} {\bibinfo  {journal}
  {Nature Nanotechnology}\ }\textbf {\bibinfo {volume} {12}},\ \bibinfo {pages}
  {1} (\bibinfo {year} {2016})}\BibitemShut {NoStop}%
\bibitem [{\citenamefont {Clevenson}\ \emph {et~al.}(2015)\citenamefont
  {Clevenson}, \citenamefont {Trusheim}, \citenamefont {Teale}, \citenamefont
  {Schr{\"{o}}der}, \citenamefont {Braje},\ and\ \citenamefont
  {Englund}}]{Clevenson2015a}%
  \BibitemOpen
  \bibfield  {author} {\bibinfo {author} {\bibfnamefont {H.}~\bibnamefont
  {Clevenson}}, \bibinfo {author} {\bibfnamefont {M.~E.}\ \bibnamefont
  {Trusheim}}, \bibinfo {author} {\bibfnamefont {C.}~\bibnamefont {Teale}},
  \bibinfo {author} {\bibfnamefont {T.}~\bibnamefont {Schr{\"{o}}der}},
  \bibinfo {author} {\bibfnamefont {D.}~\bibnamefont {Braje}}, \ and\ \bibinfo
  {author} {\bibfnamefont {D.}~\bibnamefont {Englund}},\ }\href {\doibase
  10.1038/nphys3291} {\bibfield  {journal} {\bibinfo  {journal} {Nature
  Physics}\ }\textbf {\bibinfo {volume} {11}},\ \bibinfo {pages} {393}
  (\bibinfo {year} {2015})}\BibitemShut {NoStop}%
\bibitem [{\citenamefont {Ahmadi}\ \emph {et~al.}(2017)\citenamefont {Ahmadi},
  \citenamefont {El-Ella}, \citenamefont {Hansen}, \citenamefont {Huck},\ and\
  \citenamefont {Andersen}}]{Sepehr2017}%
  \BibitemOpen
  \bibfield  {author} {\bibinfo {author} {\bibfnamefont {S.}~\bibnamefont
  {Ahmadi}}, \bibinfo {author} {\bibfnamefont {H.~A.}\ \bibnamefont {El-Ella}},
  \bibinfo {author} {\bibfnamefont {J.~O.}\ \bibnamefont {Hansen}}, \bibinfo
  {author} {\bibfnamefont {A.}~\bibnamefont {Huck}}, \ and\ \bibinfo {author}
  {\bibfnamefont {U.~L.}\ \bibnamefont {Andersen}},\ }\href {\doibase
  10.1103/PhysRevApplied.8.034001} {\bibfield  {journal} {\bibinfo  {journal}
  {Physical Review Applied}\ }\textbf {\bibinfo {volume} {8}},\ \bibinfo
  {pages} {034001} (\bibinfo {year} {2017})}\BibitemShut {NoStop}%
\bibitem [{\citenamefont {Taylor}\ \emph {et~al.}(2008)\citenamefont {Taylor},
  \citenamefont {Cappellaro}, \citenamefont {Childress}, \citenamefont {Jiang},
  \citenamefont {Budker}, \citenamefont {Hemmer}, \citenamefont {Yacoby},
  \citenamefont {Walsworth},\ and\ \citenamefont {Lukin}}]{Taylor2008}%
  \BibitemOpen
  \bibfield  {author} {\bibinfo {author} {\bibfnamefont {J.~M.}\ \bibnamefont
  {Taylor}}, \bibinfo {author} {\bibfnamefont {P.}~\bibnamefont {Cappellaro}},
  \bibinfo {author} {\bibfnamefont {L.}~\bibnamefont {Childress}}, \bibinfo
  {author} {\bibfnamefont {L.}~\bibnamefont {Jiang}}, \bibinfo {author}
  {\bibfnamefont {D.}~\bibnamefont {Budker}}, \bibinfo {author} {\bibfnamefont
  {P.~R.}\ \bibnamefont {Hemmer}}, \bibinfo {author} {\bibfnamefont
  {A.}~\bibnamefont {Yacoby}}, \bibinfo {author} {\bibfnamefont
  {R.}~\bibnamefont {Walsworth}}, \ and\ \bibinfo {author} {\bibfnamefont
  {M.~D.}\ \bibnamefont {Lukin}},\ }\href {\doibase 10.1038/nphys1075}
  {\bibfield  {journal} {\bibinfo  {journal} {Nature Physics}\ }\textbf
  {\bibinfo {volume} {4}},\ \bibinfo {pages} {810} (\bibinfo {year}
  {2008})}\BibitemShut {NoStop}%
\bibitem [{\citenamefont {Hadden}\ \emph {et~al.}(2010)\citenamefont {Hadden},
  \citenamefont {Harrison}, \citenamefont {Stanley-Clarke}, \citenamefont
  {Marseglia}, \citenamefont {Ho}, \citenamefont {Patton}, \citenamefont
  {O'Brien},\ and\ \citenamefont {Rarity}}]{Hadden2010}%
  \BibitemOpen
  \bibfield  {author} {\bibinfo {author} {\bibfnamefont {J.~P.}\ \bibnamefont
  {Hadden}}, \bibinfo {author} {\bibfnamefont {J.~P.}\ \bibnamefont
  {Harrison}}, \bibinfo {author} {\bibfnamefont {A.~C.}\ \bibnamefont
  {Stanley-Clarke}}, \bibinfo {author} {\bibfnamefont {L.}~\bibnamefont
  {Marseglia}}, \bibinfo {author} {\bibfnamefont {Y.~L.~D.}\ \bibnamefont
  {Ho}}, \bibinfo {author} {\bibfnamefont {B.~R.}\ \bibnamefont {Patton}},
  \bibinfo {author} {\bibfnamefont {J.~L.}\ \bibnamefont {O'Brien}}, \ and\
  \bibinfo {author} {\bibfnamefont {J.~G.}\ \bibnamefont {Rarity}},\ }\href
  {\doibase 10.1063/1.3519847} {\bibfield  {journal} {\bibinfo  {journal}
  {Applied Physics Letters}\ }\textbf {\bibinfo {volume} {97}},\ \bibinfo
  {pages} {241901} (\bibinfo {year} {2010})}\BibitemShut {NoStop}%
\bibitem [{\citenamefont {{Le Sage}}\ \emph {et~al.}(2012)\citenamefont {{Le
  Sage}}, \citenamefont {Pham}, \citenamefont {Bar-Gill}, \citenamefont
  {Belthangady}, \citenamefont {Lukin}, \citenamefont {Yacoby},\ and\
  \citenamefont {Walsworth}}]{LeSage2012}%
  \BibitemOpen
  \bibfield  {author} {\bibinfo {author} {\bibfnamefont {D.}~\bibnamefont {{Le
  Sage}}}, \bibinfo {author} {\bibfnamefont {L.~M.}\ \bibnamefont {Pham}},
  \bibinfo {author} {\bibfnamefont {N.}~\bibnamefont {Bar-Gill}}, \bibinfo
  {author} {\bibfnamefont {C.}~\bibnamefont {Belthangady}}, \bibinfo {author}
  {\bibfnamefont {M.~D.}\ \bibnamefont {Lukin}}, \bibinfo {author}
  {\bibfnamefont {A.}~\bibnamefont {Yacoby}}, \ and\ \bibinfo {author}
  {\bibfnamefont {R.~L.}\ \bibnamefont {Walsworth}},\ }\href {\doibase
  10.1103/PhysRevB.85.121202} {\bibfield  {journal} {\bibinfo  {journal}
  {Physical Review B}\ }\textbf {\bibinfo {volume} {85}},\ \bibinfo {pages}
  {121202} (\bibinfo {year} {2012})}\BibitemShut {NoStop}%
\bibitem [{\citenamefont {Israelsen}\ \emph {et~al.}(2014)\citenamefont
  {Israelsen}, \citenamefont {Kumar}, \citenamefont {Tawfieq}, \citenamefont
  {Neergaard-Nielsen}, \citenamefont {Huck},\ and\ \citenamefont
  {Andersen}}]{Israelsen2014}%
  \BibitemOpen
  \bibfield  {author} {\bibinfo {author} {\bibfnamefont {N.~M.}\ \bibnamefont
  {Israelsen}}, \bibinfo {author} {\bibfnamefont {S.}~\bibnamefont {Kumar}},
  \bibinfo {author} {\bibfnamefont {M.}~\bibnamefont {Tawfieq}}, \bibinfo
  {author} {\bibfnamefont {J.~S.}\ \bibnamefont {Neergaard-Nielsen}}, \bibinfo
  {author} {\bibfnamefont {A.}~\bibnamefont {Huck}}, \ and\ \bibinfo {author}
  {\bibfnamefont {U.~L.}\ \bibnamefont {Andersen}},\ }\href {\doibase
  10.1088/2040-8978/16/11/114017} {\bibfield  {journal} {\bibinfo  {journal}
  {Journal of Optics}\ }\textbf {\bibinfo {volume} {16}},\ \bibinfo {pages}
  {114017} (\bibinfo {year} {2014})}\BibitemShut {NoStop}%
\bibitem [{\citenamefont {Riedel}\ \emph {et~al.}(2014)\citenamefont {Riedel},
  \citenamefont {Rohner}, \citenamefont {Ganzhorn}, \citenamefont {Kaldewey},
  \citenamefont {Appel}, \citenamefont {Neu}, \citenamefont {Warburton},\ and\
  \citenamefont {Maletinsky}}]{Riedel2014}%
  \BibitemOpen
  \bibfield  {author} {\bibinfo {author} {\bibfnamefont {D.}~\bibnamefont
  {Riedel}}, \bibinfo {author} {\bibfnamefont {D.}~\bibnamefont {Rohner}},
  \bibinfo {author} {\bibfnamefont {M.}~\bibnamefont {Ganzhorn}}, \bibinfo
  {author} {\bibfnamefont {T.}~\bibnamefont {Kaldewey}}, \bibinfo {author}
  {\bibfnamefont {P.}~\bibnamefont {Appel}}, \bibinfo {author} {\bibfnamefont
  {E.}~\bibnamefont {Neu}}, \bibinfo {author} {\bibfnamefont {R.~J.}\
  \bibnamefont {Warburton}}, \ and\ \bibinfo {author} {\bibfnamefont
  {P.}~\bibnamefont {Maletinsky}},\ }\href {\doibase
  10.1103/PhysRevApplied.2.064011} {\bibfield  {journal} {\bibinfo  {journal}
  {Physical Review Applied}\ }\textbf {\bibinfo {volume} {2}},\ \bibinfo
  {pages} {064011} (\bibinfo {year} {2014})}\BibitemShut {NoStop}%
\bibitem [{\citenamefont {Momenzadeh}\ \emph {et~al.}(2015)\citenamefont
  {Momenzadeh}, \citenamefont {St{\"{o}}hr}, \citenamefont {{De Oliveira}},
  \citenamefont {Brunner}, \citenamefont {Denisenko}, \citenamefont {Yang},
  \citenamefont {Reinhard},\ and\ \citenamefont {Wrachtrup}}]{Momenzadeh2015}%
  \BibitemOpen
  \bibfield  {author} {\bibinfo {author} {\bibfnamefont {S.~A.}\ \bibnamefont
  {Momenzadeh}}, \bibinfo {author} {\bibfnamefont {R.~J.}\ \bibnamefont
  {St{\"{o}}hr}}, \bibinfo {author} {\bibfnamefont {F.~F.}\ \bibnamefont {{De
  Oliveira}}}, \bibinfo {author} {\bibfnamefont {A.}~\bibnamefont {Brunner}},
  \bibinfo {author} {\bibfnamefont {A.}~\bibnamefont {Denisenko}}, \bibinfo
  {author} {\bibfnamefont {S.}~\bibnamefont {Yang}}, \bibinfo {author}
  {\bibfnamefont {F.}~\bibnamefont {Reinhard}}, \ and\ \bibinfo {author}
  {\bibfnamefont {J.}~\bibnamefont {Wrachtrup}},\ }\href {\doibase
  10.1021/nl503326t} {\bibfield  {journal} {\bibinfo  {journal} {Nano Letters}\
  }\textbf {\bibinfo {volume} {15}},\ \bibinfo {pages} {165} (\bibinfo {year}
  {2015})}\BibitemShut {NoStop}%
\bibitem [{\citenamefont {Wolf}\ \emph {et~al.}(2015)\citenamefont {Wolf},
  \citenamefont {Neumann}, \citenamefont {Nakamura}, \citenamefont {Sumiya},
  \citenamefont {Ohshima}, \citenamefont {Isoya},\ and\ \citenamefont
  {Wrachtrup}}]{Wolf2015}%
  \BibitemOpen
  \bibfield  {author} {\bibinfo {author} {\bibfnamefont {T.}~\bibnamefont
  {Wolf}}, \bibinfo {author} {\bibfnamefont {P.}~\bibnamefont {Neumann}},
  \bibinfo {author} {\bibfnamefont {K.}~\bibnamefont {Nakamura}}, \bibinfo
  {author} {\bibfnamefont {H.}~\bibnamefont {Sumiya}}, \bibinfo {author}
  {\bibfnamefont {T.}~\bibnamefont {Ohshima}}, \bibinfo {author} {\bibfnamefont
  {J.}~\bibnamefont {Isoya}}, \ and\ \bibinfo {author} {\bibfnamefont
  {J.}~\bibnamefont {Wrachtrup}},\ }\href {\doibase 10.1103/PhysRevX.5.041001}
  {\bibfield  {journal} {\bibinfo  {journal} {Physical Review X}\ }\textbf
  {\bibinfo {volume} {5}},\ \bibinfo {pages} {041001} (\bibinfo {year}
  {2015})}\BibitemShut {NoStop}%
\bibitem [{\citenamefont {Jensen}\ \emph {et~al.}(2014)\citenamefont {Jensen},
  \citenamefont {Leefer}, \citenamefont {Jarmola}, \citenamefont {Dumeige},
  \citenamefont {Acosta}, \citenamefont {Kehayias}, \citenamefont {Patton},\
  and\ \citenamefont {Budker}}]{Jensen2014}%
  \BibitemOpen
  \bibfield  {author} {\bibinfo {author} {\bibfnamefont {K.}~\bibnamefont
  {Jensen}}, \bibinfo {author} {\bibfnamefont {N.}~\bibnamefont {Leefer}},
  \bibinfo {author} {\bibfnamefont {A.}~\bibnamefont {Jarmola}}, \bibinfo
  {author} {\bibfnamefont {Y.}~\bibnamefont {Dumeige}}, \bibinfo {author}
  {\bibfnamefont {V.~M.}\ \bibnamefont {Acosta}}, \bibinfo {author}
  {\bibfnamefont {P.}~\bibnamefont {Kehayias}}, \bibinfo {author}
  {\bibfnamefont {B.}~\bibnamefont {Patton}}, \ and\ \bibinfo {author}
  {\bibfnamefont {D.}~\bibnamefont {Budker}},\ }\href {\doibase
  10.1103/PhysRevLett.112.160802} {\bibfield  {journal} {\bibinfo  {journal}
  {Physical Review Letters}\ }\textbf {\bibinfo {volume} {112}},\ \bibinfo
  {pages} {160802} (\bibinfo {year} {2014})}\BibitemShut {NoStop}%
\bibitem [{\citenamefont {Wickenbrock}\ \emph {et~al.}(2016)\citenamefont
  {Wickenbrock}, \citenamefont {Zheng}, \citenamefont {Bougas}, \citenamefont
  {Leefer}, \citenamefont {Afach}, \citenamefont {Jarmola}, \citenamefont
  {Acosta},\ and\ \citenamefont {Budker}}]{Wickenbrock2016}%
  \BibitemOpen
  \bibfield  {author} {\bibinfo {author} {\bibfnamefont {A.}~\bibnamefont
  {Wickenbrock}}, \bibinfo {author} {\bibfnamefont {H.}~\bibnamefont {Zheng}},
  \bibinfo {author} {\bibfnamefont {L.}~\bibnamefont {Bougas}}, \bibinfo
  {author} {\bibfnamefont {N.}~\bibnamefont {Leefer}}, \bibinfo {author}
  {\bibfnamefont {S.}~\bibnamefont {Afach}}, \bibinfo {author} {\bibfnamefont
  {A.}~\bibnamefont {Jarmola}}, \bibinfo {author} {\bibfnamefont {V.~M.}\
  \bibnamefont {Acosta}}, \ and\ \bibinfo {author} {\bibfnamefont
  {D.}~\bibnamefont {Budker}},\ }\href {\doibase 10.1063/1.4960171} {\bibfield
  {journal} {\bibinfo  {journal} {Applied Physics Letters}\ }\textbf {\bibinfo
  {volume} {109}},\ \bibinfo {pages} {053505} (\bibinfo {year}
  {2016})}\BibitemShut {NoStop}%
\bibitem [{\citenamefont {Huxter}\ \emph {et~al.}(2013)\citenamefont {Huxter},
  \citenamefont {Oliver}, \citenamefont {Budker},\ and\ \citenamefont
  {Fleming}}]{Huxter2013}%
  \BibitemOpen
  \bibfield  {author} {\bibinfo {author} {\bibfnamefont {V.~M.}\ \bibnamefont
  {Huxter}}, \bibinfo {author} {\bibfnamefont {T.~a.~a.}\ \bibnamefont
  {Oliver}}, \bibinfo {author} {\bibfnamefont {D.}~\bibnamefont {Budker}}, \
  and\ \bibinfo {author} {\bibfnamefont {G.~R.}\ \bibnamefont {Fleming}},\
  }\href {\doibase 10.1038/nphys2753} {\bibfield  {journal} {\bibinfo
  {journal} {Nature Physics}\ }\textbf {\bibinfo {volume} {9}},\ \bibinfo
  {pages} {744} (\bibinfo {year} {2013})}\BibitemShut {NoStop}%
\bibitem [{\citenamefont {Wee}\ \emph {et~al.}(2007)\citenamefont {Wee},
  \citenamefont {Tzeng}, \citenamefont {Han}, \citenamefont {Chang},
  \citenamefont {Fann}, \citenamefont {Hsu}, \citenamefont {Chen},\ and\
  \citenamefont {Yu}}]{Wee2007}%
  \BibitemOpen
  \bibfield  {author} {\bibinfo {author} {\bibfnamefont {T.~L.}\ \bibnamefont
  {Wee}}, \bibinfo {author} {\bibfnamefont {Y.~K.}\ \bibnamefont {Tzeng}},
  \bibinfo {author} {\bibfnamefont {C.~C.}\ \bibnamefont {Han}}, \bibinfo
  {author} {\bibfnamefont {H.~C.}\ \bibnamefont {Chang}}, \bibinfo {author}
  {\bibfnamefont {W.}~\bibnamefont {Fann}}, \bibinfo {author} {\bibfnamefont
  {J.~H.}\ \bibnamefont {Hsu}}, \bibinfo {author} {\bibfnamefont {K.~M.}\
  \bibnamefont {Chen}}, \ and\ \bibinfo {author} {\bibfnamefont {E.~C.}\
  \bibnamefont {Yu}},\ }\href {\doibase 10.1021/jp073938o} {\bibfield
  {journal} {\bibinfo  {journal} {Journal of Physical Chemistry A}\ }\textbf
  {\bibinfo {volume} {111}},\ \bibinfo {pages} {9379} (\bibinfo {year}
  {2007})}\BibitemShut {NoStop}%
\bibitem [{\citenamefont {Smeltzer}\ \emph {et~al.}(2009)\citenamefont
  {Smeltzer}, \citenamefont {McIntyre},\ and\ \citenamefont
  {Childress}}]{Smeltzer2009}%
  \BibitemOpen
  \bibfield  {author} {\bibinfo {author} {\bibfnamefont {B.}~\bibnamefont
  {Smeltzer}}, \bibinfo {author} {\bibfnamefont {J.}~\bibnamefont {McIntyre}},
  \ and\ \bibinfo {author} {\bibfnamefont {L.}~\bibnamefont {Childress}},\
  }\href {\doibase 10.1103/PhysRevA.80.050302} {\bibfield  {journal} {\bibinfo
  {journal} {Physical Review A}\ }\textbf {\bibinfo {volume} {80}},\ \bibinfo
  {pages} {1} (\bibinfo {year} {2009})}\BibitemShut {NoStop}%
\bibitem [{\citenamefont {El-Ella}\ \emph {et~al.}(2017)\citenamefont
  {El-Ella}, \citenamefont {Ahmadi}, \citenamefont {Wojciechowski},
  \citenamefont {Huck},\ and\ \citenamefont {Andersen}}]{Haitham2017}%
  \BibitemOpen
  \bibfield  {author} {\bibinfo {author} {\bibfnamefont {H.~A.~R.}\
  \bibnamefont {El-Ella}}, \bibinfo {author} {\bibfnamefont {S.}~\bibnamefont
  {Ahmadi}}, \bibinfo {author} {\bibfnamefont {A.~M.}\ \bibnamefont
  {Wojciechowski}}, \bibinfo {author} {\bibfnamefont {A.}~\bibnamefont {Huck}},
  \ and\ \bibinfo {author} {\bibfnamefont {U.~L.}\ \bibnamefont {Andersen}},\
  }\href {\doibase 10.1364/OE.25.014809} {\bibfield  {journal} {\bibinfo
  {journal} {Optics Express}\ }\textbf {\bibinfo {volume} {25}},\ \bibinfo
  {pages} {14809} (\bibinfo {year} {2017})}\BibitemShut {NoStop}%
\bibitem [{\citenamefont {Robledo}\ \emph {et~al.}(2011)\citenamefont
  {Robledo}, \citenamefont {Bernien}, \citenamefont {Sar},\ and\ \citenamefont
  {Hanson}}]{Robledo2011a}%
  \BibitemOpen
  \bibfield  {author} {\bibinfo {author} {\bibfnamefont {L.}~\bibnamefont
  {Robledo}}, \bibinfo {author} {\bibfnamefont {H.}~\bibnamefont {Bernien}},
  \bibinfo {author} {\bibfnamefont {T.~V.~D.}\ \bibnamefont {Sar}}, \ and\
  \bibinfo {author} {\bibfnamefont {R.}~\bibnamefont {Hanson}},\ }\href@noop {}
  {\bibfield  {journal} {\bibinfo  {journal} {New Journal of Physics}\ }\textbf
  {\bibinfo {volume} {13}} (\bibinfo {year} {2011})}\BibitemShut {NoStop}%
\bibitem [{\citenamefont {Acosta}\ \emph {et~al.}(2010)\citenamefont {Acosta},
  \citenamefont {Jarmola}, \citenamefont {Bauch},\ and\ \citenamefont
  {Budker}}]{Acosta2010a}%
  \BibitemOpen
  \bibfield  {author} {\bibinfo {author} {\bibfnamefont {V.~M.}\ \bibnamefont
  {Acosta}}, \bibinfo {author} {\bibfnamefont {A.}~\bibnamefont {Jarmola}},
  \bibinfo {author} {\bibfnamefont {E.}~\bibnamefont {Bauch}}, \ and\ \bibinfo
  {author} {\bibfnamefont {D.}~\bibnamefont {Budker}},\ }\href {\doibase
  10.1103/PhysRevB.82.201202} {\bibfield  {journal} {\bibinfo  {journal}
  {Physical Review B}\ }\textbf {\bibinfo {volume} {82}},\ \bibinfo {pages}
  {201202} (\bibinfo {year} {2010})}\BibitemShut {NoStop}%
\bibitem [{\citenamefont {Jensen}\ \emph {et~al.}(2013)\citenamefont {Jensen},
  \citenamefont {Acosta}, \citenamefont {Jarmola},\ and\ \citenamefont
  {Budker}}]{Jensen2013}%
  \BibitemOpen
  \bibfield  {author} {\bibinfo {author} {\bibfnamefont {K.}~\bibnamefont
  {Jensen}}, \bibinfo {author} {\bibfnamefont {V.~M.}\ \bibnamefont {Acosta}},
  \bibinfo {author} {\bibfnamefont {A.}~\bibnamefont {Jarmola}}, \ and\
  \bibinfo {author} {\bibfnamefont {D.}~\bibnamefont {Budker}},\ }\href
  {\doibase 10.1103/PhysRevB.87.014115} {\bibfield  {journal} {\bibinfo
  {journal} {Physical Review B}\ }\textbf {\bibinfo {volume} {87}},\ \bibinfo
  {pages} {014115} (\bibinfo {year} {2013})}\BibitemShut {NoStop}%
\end{thebibliography}%

\end{document}